\documentclass[12pt]{article}
\usepackage{graphicx}
\usepackage{multirow}
\usepackage{subfig}
\usepackage{xspace}
\usepackage{xcolor}
\usepackage[countmax]{subfloat}
\usepackage{amsmath}
\usepackage{mathtools}
\usepackage{graphicx}
\usepackage{slashed}
\usepackage{tikz-feynman,contour}
\usepackage{tikz,pgf}
\usepackage{braket}
\usepackage{amsfonts}
\usepackage{amsthm,scrextend,bbold}
\usepackage{comment}
\newcommand{\as}{\alpha_\text{s}}
\newcommand{\cf}{C_{\text{F}}}

\newcommand{\order}[1]{{\cal O}\left(#1\right)}

\DeclareMathOperator{\De}{d}
\newcommand{\de}{\De\!}
\newcommand{\xf}{x}

\newcommand{\gszero}{\gamma_\text{soft}^{(0)}}

\newcommand{\gszerotilde}{\widetilde{\gamma}_\text{soft}^{(0)}}

\newcommand{\muf}{\mu_{\text{F}}}

\newcommand\pubdate{\today}

\def\Title#1{\begin{center} {Soft resummation in processes with heavy quark: bridging the gap from 4-flavor to 5-flavor scheme} \end{center}}
\def\Author#1{\begin{center}{ Andrea Ghira} \end{center}}
\def\Address#1{\begin{center}{Dipartimento di Fisica, Universit\`a di Genova and INFN, Sezione di Genova,Via Dodecaneso 33, 16146, Italy} \end{center}}

\newcommand\pubblock{\rightline{\begin{tabular}{l}  \\ 
         \pubdate  \end{tabular}}}
\newenvironment{Abstract}{}{}
\newenvironment{Presented}{\begin{quotation} \begin{center} 
             PRESENTED AT\end{center}\bigskip 
      \begin{center}\begin{large}}{\end{large}\end{center} \end{quotation}}

\begin{document}
\begin{titlepage}
 \pubblock
\vfill
\Title{}
\vfill
\Author{Andrea Ghira}
\Address{Dipartimento di Fisica, Universit\`a di Genova and INFN, Sezione di Genova,Via Dodecaneso 33, 16146, Italy}
\vfill
\begin{Abstract}
In this work we present a new approach to threshold resummation in processes with heavy quarks. In particular we will focus on the differential decay rate of a color-singlet particle into a $b \bar b$ pair and we will show how to resum in a consistent way both the logarithms of the mass and the logarithms of the heavy flavor energy fraction. Within this framework, we match the two different approaches existing in literature to threshold resummation, the main difference of which is the way in which the mass is treated (5-flavor scheme vs 4-flavor scheme).
\end{Abstract}
\vfill
\begin{Presented}
DIS2023: XXX International Workshop on Deep-Inelastic Scattering and
Related Subjects, \\
Michigan State University, USA, 27-31 March 2023 \\
     \includegraphics[width=9cm]{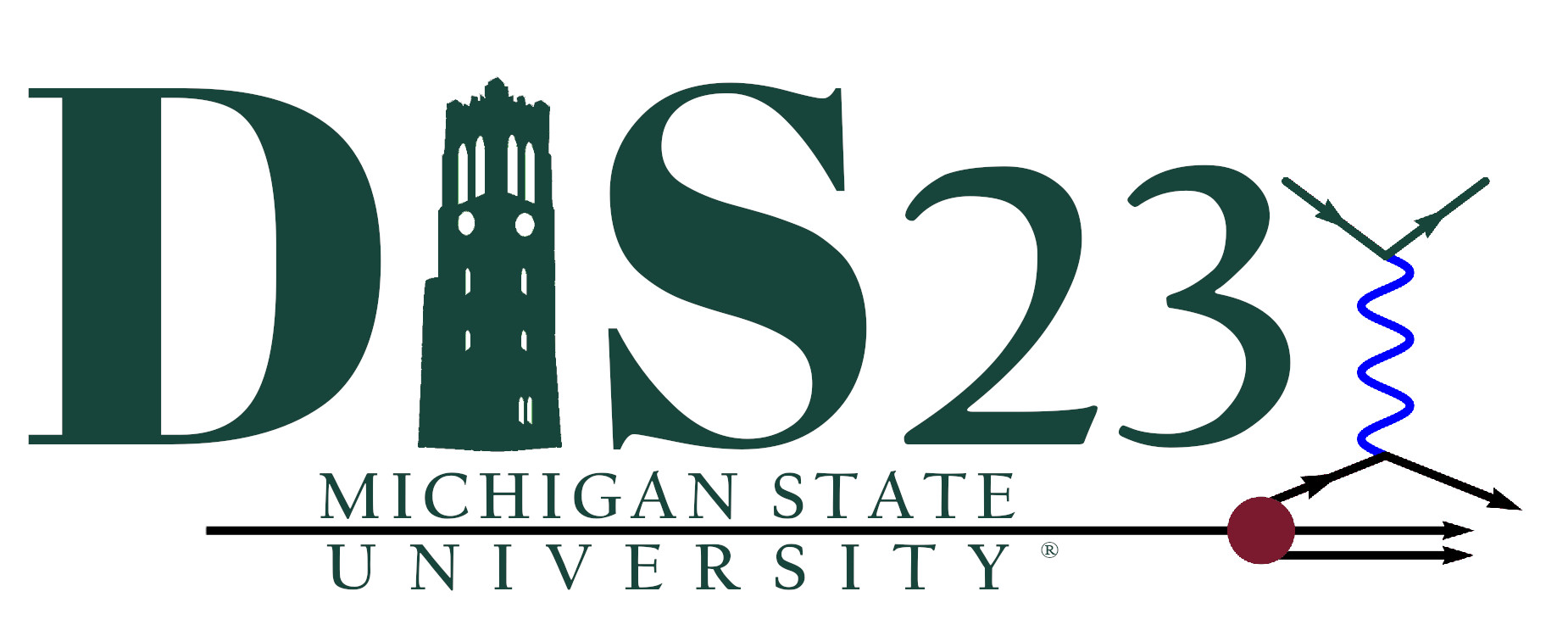}
\end{Presented}
\vfill
\end{titlepage}
\section{Introduction}
We consider the production of a $b\bar b$ pair from the decay of a color-singlet particle which can be a Higgs or $Z$ boson,
plus undetected radiation:
\begin{equation}
	h(q)\to b(p_1)+\bar{b}(p_2)+X(k)
\end{equation}
(four-momenta are indicated in brackets).
We are interested in the differential decay rate $\frac{\de \Gamma}{\de x}$ with respect to the dimensionless variable
\begin{equation}
	x=\frac{2 p_1\cdot q}{q^2},
\end{equation}
which coincides with the fraction of the total available energy carried away by the $b$ quark in the centre of mass frame. In the threshold limit, which means either soft or collinear, $x\to 1$.
In order to perform the calculation of the decay rate $\frac{\de \Gamma}{\de x}$ two main approaches are mainly employed: we can refer to the first as the massive-scheme approach (or 4-flavor) and to the second as the massless one (or 5-flavor).
In the first framework, the spectrum $\frac{\de \Gamma}{\de x}$ is computed, up to some finite order in perturbation theory, taking into exact account the finite value of the heavy quark mass $m$. Collinear singularities are regularized by the heavy quark mass and therefore $\log\frac{m^2}{q^2}$ appear in the perturbative coefficients: such logarithms may eventually spoil the convergence of the series. On the other hand, the kinematics of radiation is treated correctly at every perturbative order. 
In the second approach, the heavy flavour mass is only used as a regulator of collinear divergences, while contributions proportional to powers of $\frac{m^2}{q^2}$ are systematically neglected. This means that we do not have control on the kinematics of the emission outside the collinear region.
This framework exploits a factorization theorem: the decay rate is written as a convolution of process independent fragmentation functions $\mathcal{D}_i$ and of process-dependent partonic cross-sections: 
\begin{align}\label{eq:rate-frag}
	\frac{1}{\Gamma_0}\frac{\de\Gamma}{\de\xf}&=  \sum_{i} \int_x^1 \frac{\de z}{z} \mathcal{C}_i\left(\frac{x}{z},\as, \frac{\mu^2}{q^2} \right) \mathcal{D}_i (z,\mu^2,m^2) +\order{\frac{m^2}{q^2}},
\end{align}
Here $\mu^2$ is the factorization scale typically chosen oh the order of $q^2$ and $\Gamma_0$ is the decay rate at Born level. The sum runs over all the possible partons that can fragment in the heavy quark.
Logarithms of the mass can be resummed to all orders up to a given logarithmic accuracy thanks to DGLAP evolution equation,
\begin{align}\label{eq:dglap}
	\mu^2 \frac{\de }{\de \, \mu^2}\widetilde{\mathcal{D}}_i (N,\mu^2,m^2) =& \sum_j \gamma_{ij}\left(N,\as(\mu^2) \right) \widetilde{\mathcal{D}}^j (N,\mu^2,m^2),\\
	\implies \widetilde{\mathcal{D}}_i (N,\mu^2,m^2)=& \mathcal{E}_{ij}(N,\mu^2,\mu_0^2) \widetilde{\mathcal{D}}_0(\mu_0^2,m^2),
\end{align}

where $\widetilde{\mathcal{D}}^j$ denotes the Mellin transform of the $j$-th fragmetation function defined as:
\begin{equation}
	\widetilde{f}(N)=\int^1_0 \de x \; x^{N-1} f(x).
\end{equation}
for some function $f$. $\gamma_{ij}$ are the Mellin transform of the Altarelli Parisi splitting function, while $\mathcal{E}$ is the so called DGLAP evolution operator.
The initial condition of the fragmentation functions $\widetilde{\mathcal{D}}_0$ at a given scale $\mu_0^2$ is needed to completely solve the differential equation and for the $b$ quark we can naturally set this scale such that $\mu_0^2\simeq m^2$. This means that the initial condition can be computed perturbatively~\cite{Mele:1990cw,Cacciari:2001cw, Melnikov:2004bm}.
Merging these two different frameworks gives us a better prediction for the different regions of $q^2$
\begin{equation}\label{FONLL}
	\widetilde{\Gamma}(N,\xi)= \widetilde{\Gamma}_k^{(4)}(N,\xi)+\widetilde{\Gamma}^{(5)}_\ell(N,\xi)-\text{double counting}, \quad \xi=\frac{m^2}{q^2}.
\end{equation}
with $\widetilde{\Gamma}^{(4)}(N,\xi)$ the Mellin transform of the massive calculation at perturbative order $k$  and $\widetilde{\Gamma}^{(5)}_\ell(N,\xi)$ is the Mellin transform of the massless decay rate at logarithmic accuracy $\ell$.
Finally the ``double counting" is the perturbative expansion of  $\widetilde{\Gamma}^{(5)}_\ell(N,\xi)$ to order $k$. We will restrict ourselves to the case $\ell=1$ (FONLL)~\cite{Cacciari:1998it}.
Our main task is to generalize FONLL scheme including also the threshold resummation. In the next section we now explain the main difficulties in the merging of the two approaches.
\section{Logarithmic structure at large $N$}
Both the quantities appearing in the rhs of Eq.~(\ref{FONLL}) display a logarithmically divergent behaviour as $N\to\infty$ due to the presence in the coefficients of the physical spectrum $\frac{\de \Gamma}{\de x}$ of distributions:
\begin{equation}
	d_k(x)=\left[\frac{\log^{k-1}(1-x)}{1-x}\right]_+
	\label{dk}
\end{equation}
 which diverge at the threshold $x\to 1$.
 The limit $x\to 1$ is mapped into $N\to \infty$  by the Mellin transform. 
These logarithmic contributions can be resummed to all orders up to a given logarithmic accuracy but the merging of the two resummed formula is far from trivial 
due to the different structures at large $N$. 
Specifically, in the large $N$ limit $\widetilde{\Gamma}^{(4)}_k(N,\xi)$ is a polynomial of degree $n$ in $\log N$ whose coefficients $c^{(4)}_n(\xi)$ carry the full $\xi$ dependence:
\begin{equation}
	\widetilde{\Gamma}^{(4)}_k(N,\xi)=\sum^k_{n=0} c^{(4)}_n(\xi) \left(\frac{\as}{\pi}\right)^n \log^n{N} +\mathcal{O}\left(\frac{1}{N}\right)
\end{equation}
On the other hand expanding $\widetilde{\Gamma}^{(5)}_\ell(N,\xi)$ at $\order{\as^k}$ we obtain a polynomial of degree $2n$ in $\log N$, whose coefficients depend only on the log of the mass: 
\begin{equation}
   	\widetilde{\Gamma}^{(5)}_k(N,\xi)=\sum^k_{n=0} c^{(5)}_n(\xi) \left(\frac{\as}{\pi}\right)^n \log^{2n}{N} +\mathcal{O}\left(\frac{1}{N}\right)
\end{equation}
At $\order\as$ expanding the threshold-resummed formulas in \cite{Cacciari:2001cw,Laenen:1998qw} we find:
\begin{equation} \label{eq:double limit}
	\begin{split}
		\widetilde{\Gamma}^{(5)}_{k=1}(N,\xi)=& 1+\frac{\as \cf}{\pi}\left(-\frac{1}{2}\log^2{\bar N} +\log{\xi}\log{\bar N}+\frac{7}{4} \log \bar N-\frac{3}{4}\log \xi\right)+\mathcal{O}\left(N^0\right)\\
		\widetilde{\Gamma}_{k=1}^{(4)}(N,\xi)=&1+\frac{\as\cf}{\pi}\left(\frac{1}{2}\log^2 \xi+2\log{\bar N}\log\xi+2\log \bar N-\frac{1}{2}\log \xi\right)+ \mathcal{O}\left(N^0,\xi^0\right)
	\end{split}
\end{equation}
with $\bar N= N e^{\gamma_{\text{E}}}$, $\gamma_{\text{E}}$ the Euler-Mascheroni constant.
Equation (\ref{eq:double limit}) presents two main problems: the first, as already outline, is the fact that the five flavour formalism contains double log of $N$ whereas the four flavor do not. This is a consequence of the fact that soft and small mass limit does not commute \cite{Gaggero:2022hmv,Corcella:2003ib}. 
Another important problem that cause a mismatch between the two formulas is that also the mass logs have different structures. At $\order{\as}$ within the massless framework only at most single logs of the mass are present, whereas in Eq (\ref{eq:double limit}) also double logs of the mass appear \cite{Gaggero:2022hmv}.
Due to the non-commutativity of the limits it is impossible to define a matching scheme like FONLL: the main problem is that we do not know how to identify an all order double counting term.
In the following, we will present a solution to this problem, valid to next-to-leading log accuracy.
\section{4 vs 5 flavor-scheme}
The explicit calculation of $\widetilde{\Gamma}^{(4,\text{res})}_{\ell_1}(N,\xi)$, where $\ell_1$ denotes the logarithmic accuracy of the threshold resummation, was performed in Ref.~\cite{Laenen:1998qw}. In particular for the case $\ell_1=0$:
\begin{align}\label{eq:massive resummation}
	\widetilde{\Gamma}^{(4,\text{res})}_{\ell_1=0}(N,\xi)=\left(1+\frac{\as \cf}{\pi}\mathcal{K}^{(1)}(\xi,\as)\right) e^{-2\int^1_{1/\bar N} \frac{\de z}{z} \as\left(z^2 \mu_0^2\right) \gszero\left(\xi\right)},
\end{align}
with $\mathcal{K}^{(1)}(\xi,\as)$ a process-dependent factor that exhibits the double mass log in the massless limit, and $\gszero$, the so-called first order massive soft anomalous dimension:
\begin{equation}
	\gszero(\xi)=\cf\left(\frac{1+\beta^2}{2\beta}\log\frac{1+\beta}{1-\beta}-1\right), \quad \beta=\sqrt{1-4\xi}.
	\label{eq:gammazero}
\end{equation}
On the other hand in the five flavor approach the complete calculation was performed in \cite{Cacciari:2001cw} and in \cite{Corcella:2004xv} for $\ell_2=1$, where $\ell_2$ denotes the logarithmic accuracy of the threshold resummation in the 5 flavor scheme: it was shown that the resummed decay rate in Mellin space in can be seen as the product of two independent jet functions:
\begin{align}\label{eq:FF-sep}
	\widetilde{\Gamma}_{\ell=1, \ell_2=1}^{(5,\text{res})}(N,\xi)&=
	\left(1+ \frac{\as(\mu^2) \cf}{\pi}\mathcal{C}_0^{(1)} \right) \left(1+ \frac{\as(\mu_0^2) \cf}{\pi}\mathcal{D}_0^{(1)} \right) \tilde{\mathcal{E}}(N,\mu_0^2,\mu^2,\as(\muf^2))  \nonumber\\ &
	\exp \left[J\left(N,\frac{\mu^2}{q^2},\as(\mu^2),\frac{\mu_0^2}{m^2},\as(\mu_0^2)\right)+\bar{J}\left(N,\frac{\mu^2}{q^2},\as(\mu^2)\right)\ \right].
\end{align}
The factor $\exp(J)$ in Eq.~(\ref{eq:FF-sep}) describes  soft radiation emitted collinearly to the tagged $b$-quark and has the following form: 
\begin{equation}\label{eq:j-def}
	J=D_0 + E+\Delta,
\end{equation}
where
\begin{equation}\label{eq:evolution operator integral}
	E=-\int^{\mu^2}_{\mu_0^2} \frac{\de k^2}{k^2}  \left\{A(\as(k^2)) \log \bar{N} + \frac{1}{2}B(\as(k^2)) \right\},
\end{equation}
is the logarithmically enhanced contribution in the large $N$ limit to the DGLAP evolution kernel and
\begin{align}\label{eq: delta function}
	\Delta=&\int_{\frac{1}{\bar{N}}}^1 \frac{\de z}{z}\int^{\mu^2}_{z^2q^2}\frac{\de k^2}{k^2} A(\as(k^2)),\\
	\label{eq: initial condition}
	D_0=& -\int_{\frac{1}{\bar{N}}}^1  \frac{\de z}{z} \left\{ \int^{\mu_0^2}_{z^2 m^2} \frac{\de k^2}{k^2} A(\as(k^2))+H\left(\as\left(z^2 m^2\right)\right)\right\}.
\end{align}
On the other hand $\bar J$ describe, collinear radiation with respect to the anti-quark.
\begin{align}
	\label{eq: barJ}
	\bar{J}=& -\int_{\frac{1}{\bar{N}}}^1  \frac{\de z}{z} \left\{\int^{zq^2}_{z^2q^2}\frac{\de k^2}{k^2} A(\as(k^2))+\frac{1}{2}B(\as\left(zq^2\right))\right\}.
\end{align}
The remaining terms in Eq. (\ref{eq:FF-sep}) are costant or vanish by construction in the large $N$ limit.
\begin{figure}
	\centering
	\includegraphics[width=0.95\textwidth,page=1]{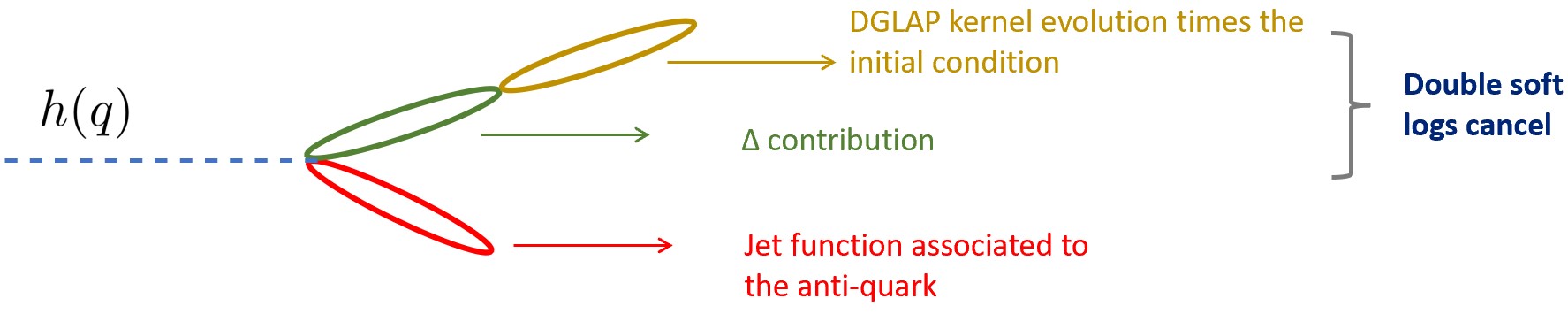}
	\caption{Here he cancellation of the double soft logs is pictorially represented: in the jet function at order $\order\as$ the double soft logs actually cancel between the initial condition $D_0$ and $\Delta$. The only remaining $\log^2{N}$ which appear in Eq (\ref{eq:double limit}) comes from $\bar J$.}
	\label{fig:jetfunctions}
\end{figure}
For sake of simplicity in the following we will perform a fixed coupling analysis.
We see from Eqs.~(\ref{eq: delta function}) and (\ref{eq: initial condition}) that the jet function $J$ does not contain any double threshold log of $N$ (see Fig.~\ref{fig:jetfunctions}). Therefore the double logs of $N$ that appears in Eq.~(\ref{eq:double limit}) has to be addressed only to the recoiling jet function $\bar J$.
We modify it in such a way that when $\frac{1}{N}>\xi$ we recover \cite{Cacciari:2001cw} and in the opposite case we recover the massive expression \cite{Laenen:1998qw} in the small mass limit.

\begin{figure}
	\centering
	\includegraphics[width=0.55\textwidth,page=1]{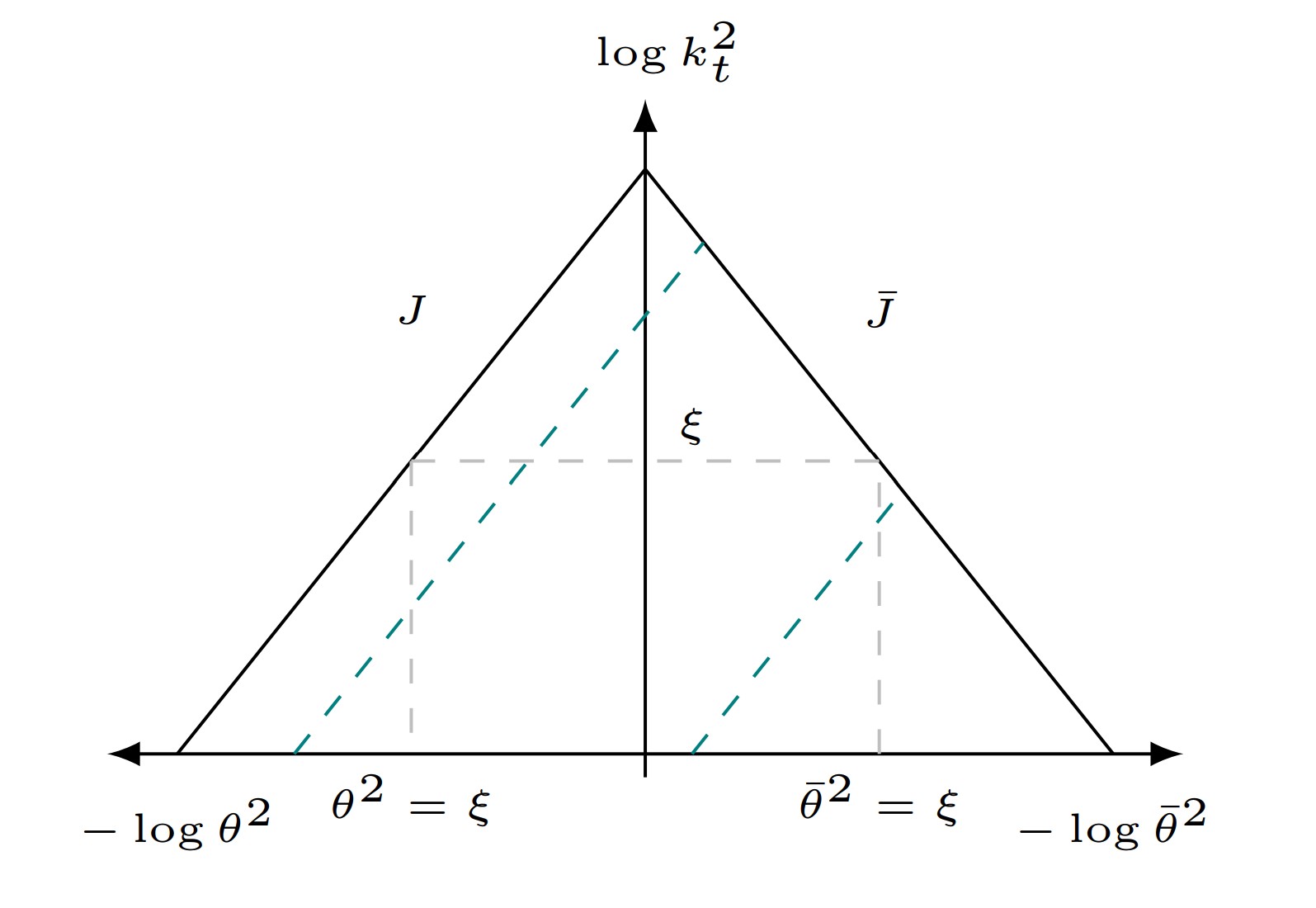}
	\caption{At this logarithmic accuracy the phase space of the emission can be visualized as two different hemispheres ($J$ and $\bar J$) and it is represented by the Lund plane in figure. Here $k_t^2$ is the transverse momentum of the emitted radiation, $\theta^2$ the emission angle off the $b$ and $\bar\theta^2$ the one off the $\bar b$. The computation of the area above the dashed teal line corresponds to the calculation of $J$ and $\bar J$. When we approach the dead-cone angle ($\bar \theta^2\simeq \xi$) mass effects have to be taken into account both for $J$ and for $\bar J$.}
	\label{LundPlane}
\end{figure}

This is achieved including finite mass effects in the computation of $\bar J$ which means that the recoil jet function has to be computed in the quasi-collinear limit. 
We outline the fact that in the 5-flavor scheme the jet function $J$ is already computed in the quasi-collinear limit ($b$ mass effects are taken into account by the initial condition $D_0$) but the $\bar b$ is assumed to be massless, therefore double logs of $N$ appear at $\order \as$.
We can visualize the meaning of this claim in the Lund plane in Fig.~(\ref{LundPlane})

By definition in the quasi collinear limit the ratio $\xi$ is kept of the same order of the emission angle off the $\bar b$ quark  $\bar \theta^2$:
\begin{align} \label{eq:jet-function-Jbar}
	\bar J(N,\xi) &=-\int_0^1 \de \bar{z}\int_0^{q^2} \frac{\de k_t^2}{k_t^2+\bar{z}^2 m^2} \frac{\as^\text{CMW}(k_t^2)}{2\pi} P_{\mathcal{Q}g} (\bar z,k_t^2)\Theta(1-\bar{\theta}^2)  \Theta\left(  \bar{z} \left(\bar \theta^2 +\xi \right) -\frac{1}{\bar N}\right).
\end{align}
where $P_{\mathcal{Q}g}$ denotes the massive splitting function:
\begin{equation}
	P_{\mathcal{Q}g}(\bar z,k_t^2)= \cf\left(\frac{1+(1-\bar z)^2}{\bar z}-\frac{2 \bar z(1-\bar z)m^2}{k_{t}^2+\bar z^2 m^2}\right),
\end{equation}
and $\bar z$ the fraction of energy taken away by the gluon from the heavy quark.
Solving the integral at fixed coupling we find:
\begin{equation} \label{eq: Jbar fix coupling}
	\begin{split}
		\text{if}\quad  \frac{1}{\bar N}> \xi, \quad \bar{J}(N,\xi)&=\frac{\as\cf}{\pi}\left(-\frac{1}{2}\log^2{\bar N}+\frac{3}{4}\log{\bar N}\right),\\
	\text{if}\quad \frac{1}{\bar N}<\xi, \quad \bar{J}(N,\xi)&=\frac{\as\cf}{\pi}\left(\frac{1}{2}\log^2{\xi}+\log{\bar N}\log{\xi}+\log{\bar N}+\frac{1}{4}\log{\xi}\right). \nonumber
		\end{split}
\end{equation}
We explicitly see that taking into account mass effects in $\bar J$ we obtain two different regimes: when $\frac{1}{\bar N}>\xi$ we get the double log of $N$ as expected. On the other hand when $\frac{1}{\bar N}<\xi$ we recover the double mass log in Eq~(\ref{eq:double limit}), meaning that the mass in the log squared is the one of the $\bar b$.
\section{Final Resummed Expression}
Once we have included the running coupling corrections we are able to obtain an all order resummed differential decay rate that interpolates consistently between the 5-flavor resummed expression and the 4-flavor one.
\begin{equation}\label{eq:final-result}
	\frac{1}{\Gamma_0}\frac{\de \Gamma}{\de x}= \int_{c-i \infty}^{c+ i \infty}\frac{\de N}{2 \pi i}\, x^{-N}
	\begin{cases} 	
		\widetilde{\Gamma}^{(1)}(N,\xi), & \text{if}\; 1-x >\sqrt{\xi},\\
		\widetilde{\Gamma}^{(2)}(N,\xi), & \text{if}\; \xi<1-x <\sqrt{\xi}, \\
		\widetilde{\Gamma}^{(3)}(N,\xi), & \text{if} \; 1-x <\xi,\\
	\end{cases}	
\end{equation}
with
\begin{align}\label{eq:NLL2}
	\widetilde{\Gamma}^{(1)}(N,\xi)&=\widetilde{\Gamma}_{\ell=1, \ell_2=1}^{(5,\text{res,sub})} \exp  \left[J^{(1)}+\bar{J}^{(1)}\right], \nonumber \\
	\widetilde{\Gamma}^{(2)}(N,\xi)&= \widetilde{\Gamma}^{(\text{match})} \exp \left[J^{(2)}+\bar{J}^{(2)}\right], \nonumber \\
	\widetilde{\Gamma}^{(3)}(N,\xi)&= \widetilde{\Gamma}^{(4,\text{res,sub})}_{\ell_1=0}  \exp \left[J^{(2)}+\bar{J}^{(3)}\right].
\end{align}
$\widetilde{\Gamma}_{\ell=1, \ell_2=1}^{(5,\text{res,sub})},\widetilde{\Gamma}^{(4,\text{res,sub})}_{\ell_1=0},$ are the subtracted version of the 5 and 4 flavour resummation:
the subtracted 5-flavour result is defined starting from Eq.~(\ref{eq:FF-sep}):
\begin{align}
	\widetilde{\Gamma}_{\ell=1, \ell_2=1}^{(5,\text{res-sub})}(N,\xi)&=
	\left(1+ \frac{\as(\mu^2) \cf}{\pi}\mathcal{C}_0^{(1)} \right) \left(1+ \frac{\as(\mu_0^2) \cf}{\pi}\mathcal{D}_0^{(1)} \right) \tilde{\mathcal{E}}(N,\mu_0^2,\mu^2,\as(\mu^2)),
\end{align}
By construction $\widetilde{\Gamma}^{(1)}$ coincides with the 5-flavour result of~\cite{Cacciari:2001cw}. 
Similarly, $\widetilde{\Gamma}^{(4,\text{res,sub})}_{\ell_1=0}$ is built from Eq~.(\ref{eq:massive resummation}): it is the resummed 4-flavor calculation subtracted by all the logs of $N$ and $\xi$ which are not power suppressed since they are already taken into account in the jet functions: 
\begin{align}
	\widetilde{\Gamma}^{(4,\text{res-sub})}_{\ell_1=0}(N,\xi)&= \left(1+ \frac{\as(\mu_0^2)}{\pi}\mathcal{K}^\text{sub}_1(\xi)\right) \exp \left[-2\, \gszerotilde(\beta) \int^1_{1/\bar N} \frac{\de z}{z}   \frac{\as\left(z^2 \mu_0^2\right)}{\pi}\right].
\end{align}
 $\widetilde{\Gamma}^{(3)}$ coincide with the four flavor calculation in \cite{Laenen:1998qw,Gaggero:2022hmv} with the difference that also mass logs have been resummed.
Finally $\widetilde{\Gamma}^{(\text{match})}$ is a matching function that interpolates between the two subtracted expressions.
The expression of the jet functions $J^{(i)}$ and $\bar J^{(i)}$ are derived including running coupling corrections in the decoupling scheme. 
\section{Conclusions}
We have developed a resummed expression that effectively bridges the gap between the 5-flavor scheme and the 4-flavor scheme. Specifically, when $\frac{1}{\bar N}>\xi$, our approach aligns with the results obtained in \cite{Cacciari:2001cw}, whereas when $\frac{1}{\bar N}<\xi$, we reproduce the findings of \cite{Laenen:1998qw}. This implies that the double threshold logarithms present in the massless scheme are inherently linked to the double logarithms of the mass observed in the massive framework.

Our derivation heavily relies on the NLL approximation, which allows for the separation of the resummed expression into the computation of two distinct jet functions. An interesting avenue for future research would involve extending this framework to NLL accuracy, which would require accounting for gluon interference between the hard particles.

It is important to note that these results are currently being prepared for publication and are the outcome of collaborative work with S. Marzani and G. Ridolfi \cite{Ghira:2023xv}.
 \\
 \\
We thank Simone Marzani and Giovanni Ridolfi for the aid in the drafting of this proceeding.
We thank Simone Caletti, Matteo Cardi, Samuele Grossi for useful discussions on this topic. 

\bibliography{references.bib}
\bibliographystyle{unsrt}
\end{document}